# Left-handed polarized spin waves in ferromagnets induced by spin-transfer torque

Zhen-wei Zhou[1], Xi-guang Wang[1], Yao-zhuang Nie[1], Qing-lin Xia[1], Zhong-ming Zeng[2], Guang-hua Guo[*,1]

[1]*School of Physics and Electronics, Central South University, Changsha, 410083, China*

[2]*Suzhou Institute of Nano-tech and Nano-bionics, Chinese Academy of Sciences, Suzhou, 215123, China*

**ABSTRACT:** Polarization is a fundamental property of waves that refers to the orientation of the oscillations. It has been widely used to encode information in photonics and phononics. However, the polarization of spin waves is rarely used yet in magnonics. The reason for this is that only the right-handed polarized spin waves can be accommodated in ferromagnets. Here, we report that stable left-handed polarized spin waves can be introduced into ferromagnets if a spin-polarized electrical current is presented. The right-handed and left-handed polarized spin waves coexist when the current density is larger than a critical value while the system keeps stable. The results are confirmed by micromagnetic simulations. This work provides new playgrounds to study spin waves and points to new findings for future experimental studies.

---

[*] Correspondence to: guogh@mail.csu.edu.cn

The finding of spin-transfer torques (STTs)[1,2] induced by the spin-polarized electrical current brings a great impact on spintronics[3]. STT can reverse magnetization[4-6], induce magnetization self-oscillations[7-9], and drive domain wall motion[10], which are exploited to design novel spintronic devices, such as STT-MRAM[11], STT-oscillators[12,13], domain-wall racetrack memory[14] and so on. The generation and manipulation of spin waves via STT have also attracted great interest because of its applications in magnonics[15-20]. The propagating spin wave excited by STT was observed directly in extended Permalloy film[17,18]. The integration of STT spin-wave sources with magnetic waveguides was achieved[19]. STT also leads to a series of new spin wave modes, such as the standing spin-wave bullets[21,22] and the magnetic droplets[23,24] in magnetic nanocontact. Due to the antidamping effect, STT can be used to control the attenuation and even to amplify the amplitude of propagating spin waves[25,26]. It induces the spin-wave Doppler shift[27,28]. This effect provides a way to measure the spin polarization of the current, the intrinsic Gilbert damping constant, and the coefficient of the nonadiabatic spin transfer torque[29].

In this work, we report a new STT effect, the STT-induced left-handed polarized spin wave (LPSW) in ferromagnets. Polarization is an intrinsic property of spin waves. In antiferromagnets, due to the two opposite magnetic sublattices, the spin waves are polarization degenerate, i.e. the right-handed polarized spin wave (RPSW)) and LPSW coexist[30]. However, in ferromagnets with one magnetic lattice, only the RPSW can be accommodated. For this reason, the polarization of spin waves is rarely exploited yet in magnonics, unlike the photon polarization which is widely used to encode information in photonics. Here we show that the stable LPSW can be introduced into ferromagnets if a spin-polarized electrical current larger than a critical value is presented. The critical current for inducing the LPSW can be greatly reduced in systems that involve the Dzyaloshinskii-Moriya interaction (DMI).

Consider a one-dimensional ferromagnet along the *y*-axis. The stable magnetization is directed in the positive *z*-axis by an external magnetic field. The magnetization dynamics is governed by the Landau-Lifschitz-Gilbert (LLG) equation including STT terms as follows[31]

$$\frac{\partial \mathbf{m}}{\partial t} = -\gamma \mathbf{m} \times \mathbf{H}_{eff} + \alpha \mathbf{m} \times \frac{\partial \mathbf{m}}{\partial t} + c_j \frac{\partial \mathbf{m}}{\partial y} - \beta c_j \mathbf{m} \times \frac{\partial \mathbf{m}}{\partial y} \qquad (1)$$

Where $\mathbf{m} = \mathbf{M}/M_s$ is the unit of the magnetization, $M_s$ is the saturation magnetization, $\gamma$ is the gyromagnetic ratio, and $\alpha$ is the Gilbert damping parameter. The effective field $\mathbf{H}_{eff}$ consists of exchange field, anisotropic field, and external magnetic field, $\mathbf{H}_{eff} = (2A/(\mu_0 M_s))(\partial^2 \mathbf{m}/\partial y^2) + (2K/(\mu_0 M_s))m_z \hat{\mathbf{e}}_z + H_{ext}\hat{\mathbf{e}}_z$. Here *A* is the exchange stiffness, *K* is the anisotropy constant, $\mu_0$ is the vacuum permeability, and $H_{ext}$ is the magnetic field applied along the *z*-direction. The last two terms on the right of Eq. (1) are the adiabatic and nonadiabatic STTs induced by the spin-polarized current flowing in the *y*-axis direction. The STT coefficient $c_j$ is proportional to the electrical current density $c_j = g\mu_B Pj/(2eM_s)$, where *g* is the Lande factor, $\mu_B$ is the Bohr magneton, *P* is the spin polarization of the current, *j* is the current density, and *e* is the electron charge. $\beta$ is the nonadiabatic torque parameter.

Let $\mathbf{m} = \mathbf{m}_0 + \delta\mathbf{m}$. Here $\delta\mathbf{m} = m_x \hat{\mathbf{e}}_x + m_y \hat{\mathbf{e}}_y$ is the dynamic component, and $\mathbf{m}_0$ describes the static state. By neglecting the high-order terms, one can rewrite the Eq. (1) in the linear form:

$$\frac{\partial m_x}{\partial t} + \alpha \frac{\partial m_y}{\partial t} = -\gamma H m_y + \gamma q \frac{\partial^2 m_y}{\partial y^2} + c_j \frac{\partial m_x}{\partial y} + \beta c_j \frac{\partial m_y}{\partial y}, \qquad (2)$$

$$\frac{\partial m_y}{\partial t} - \alpha \frac{\partial m_x}{\partial t} = \gamma H m_x - \gamma q \frac{\partial^2 m_x}{\partial y^2} + c_j \frac{\partial m_y}{\partial y} - \beta c_j \frac{\partial m_x}{\partial y}. \qquad (3)$$

Here $q = 2A/(\mu_0 M_s)$ and $H = 2K/(\mu_0 M_s) + H_{ext}$. The spin wave $\delta\mathbf{m}$ can be expressed as $m_{x,y} = m_{x0,y0}\exp[i(\omega t - ky)]$, where *k* and $\omega$ are the wave vector and frequency of the spin wave, respectively. Substituting this expression into Eq. (2) and (3), one finds the dispersion relations of spin waves as:

$$\begin{cases} \text{Re}\,\omega_+(k) = \gamma(H+qk^2) - c_j k \\ \text{Im}\,\omega_+(k) = \alpha\gamma(H+qk^2) - (\alpha-\beta)c_j k \end{cases}, \quad (4)$$

and

$$\begin{cases} \text{Re}\,\omega_-(k) = -\gamma(H+qk^2) - c_j k \\ \text{Im}\,\omega_-(k) = \alpha\gamma(H+qk^2) + (\alpha-\beta)c_j k \end{cases}. \quad (5)$$

Here $\text{Re}\,\omega$ and $\text{Im}\,\omega$ are the real and imaginary parts of frequency. $\omega_+$ corresponds to the solution $m_x = im_y$, representing the RPSW mode, and $\omega_-$ corresponds to the solution $m_x = -im_y$, so the spin wave is LPSW. In the absence of the spin-polarized current, the LPSW with negative frequency is equivalent to the RPSW with positive frequency. Physically, the spin waves with negative frequency do not exist, meaning that ferromagnets can accommodate only the RPSW. The presence of spin-polarized current breaks the symmetry and induces the spin-wave Doppler shift[27,28]. The dispersion curve $\text{Re}\,\omega_+(k)$ (or $\text{Re}\,\omega_-(k)$) moves towards the left or right depending on the direction of the current as shown in Fig. 1 (curve 2 and 3). The spin-polarized current also enhances or attenuates the spin waves. The spin waves become unstable when $\text{Im}\,\omega(k) \leq 0$ which gives the critical current coefficient $c_{j1}(k)$ at the onset of instability[32,33,35,36]. The minimum $c_{j1}$ can be obtained by setting $\partial c_{j1}(k)/\partial k = 0$:

$$c_{j1} = \frac{2\alpha}{|\alpha-\beta|}\gamma\sqrt{qH}. \quad (6)$$

Another effect of the spin-polarized current is that it also shifts the dispersion curve downward (or upward) and even makes curves across the horizontal ordinate as shown in Fig. 1 (curve 3). The critical current coefficient for the $\mathrm{Re}\omega_{-\mathrm{max}} \geq 0$ (or $\mathrm{Re}\omega_{+\mathrm{min}} \leq 0$) is

$$c_{j2} = 2\gamma\sqrt{qH} . \qquad (7)$$

Importantly, the $c_{j2}$ can be smaller than the minimum critical current coefficient $c_{j1}$ for the spin wave instability, i.e., the spin-polarized electrical current induces the LPSW with positive frequency while the system keeps stable. In other words, ferromagnets may accommodate not only the RPSWs but also the LPSWs if the current satisfies $c_{j2} < |c_j| < c_{j1}$. This condition requires the nonadiabatic parameter $\beta$ meeting $0 < \beta < 2\alpha$. Therefore, the nonadiabatic STT is indispensable for inducing the LPSW. If $\beta = \alpha$, the critical current coefficient $c_{j1}$ of instability is infinite[25,32,34,35], and the LPSW always exists as long as sufficiently large current is presented.

The LPSWs exist between $0 < \omega < \mathrm{Re}\omega_{-\mathrm{max}}$, and the wave vectors are limited in the range:

$$\frac{-c_j - \sqrt{c_j^2 - 4\gamma^2 qH}}{2\gamma q} < k < \frac{-c_j + \sqrt{c_j^2 - 4\gamma^2 qH}}{2\gamma q} . \qquad (8)$$

Due to the requirement of $\mathrm{Re}\omega_+ > 0$, the wave vector of the RPSW should satisfy:

$$k > \frac{c_j + \sqrt{c_j^2 - 4\gamma^2 qH}}{2\gamma q}, \quad \mathrm{or} \quad k < \frac{c_j - \sqrt{c_j^2 - 4\gamma^2 qH}}{2\gamma q} . \qquad (9)$$

Figure 2 shows the phase diagram for the RPSW and LPSW as well as the spin wave instability. Here the longitudinal coordinate is a reduced current coefficient $u = c_j/c_{j0}$, and $c_{j0} = \gamma(qH)^{1/2}$. In the area between the two straight lines described by $u = \pm 2$, only the RPSWs exist. Ferromagnets can accommodate both the RPSW and LPSW in the region between the two straight lines and the two curves defined by $u = \pm 2\alpha/|\alpha - \beta|$. Outside the two curves, spin waves lose their stability.

To confirm the validity of these findings, we perform micromagnetic simulations for a realistic situation: a 10 μm long one-dimensional permalloy magnet. Micromagnetic simulations are performed by numerically solving the Landau-Lifshitz-Gilbert equation including the spin-transfer torques (Eq. 1)). The magnet is divided into mesh with cell size of 1 nm. The parameters used for simulations are: $M_s = 8.6 \times 10^5$ A/m, $A = 1.3 \times 10^{-11}$ J/m, $K = 0$, $H_{ext} = 50$ mT, $\alpha = 0.01$, and $\beta = 0.014$. The corresponding critical current coefficient $c_{j1}$ for leading to the spin-wave instability is 1081 m/s and $c_{j2}$ for inducing the LPSWs is 432 m/s. Figure 1 shows the dispersion relations of spin waves at $c_j = 0$ (curve 1), -323 m/s (curve 2) and -808 m/s (curve 3). The spin waves are excited by a field of the form[37] $h_x(x,t) = h_{x0} \frac{\sin(2\pi f_c t)}{2\pi f_c t} \frac{\sin[k_c(y-5000)]}{k_c(y-5000)}$. Here $h_{x0} = 5$ mT, $f_c = 100$ GHz, and $k_c = 0.5$ nm$^{-1}$. For comparison, the dispersion relations calculated using Eqs. (4) and (5) are also given in Fig. 1 (dotted lines). The simulated and analytically calculated results are in full agreement. When $c_{j2} < |c_j| = 808$ m/s $< c_{j1}$, dispersion curve 3 shifts across the horizontal ordinate and LPSWs appear in the range $0 < \omega < \text{Re}\omega_{max}$ together with the RCP spin waves.

To further confirm the existence of the LPSWs, we excite the spin wave using a harmonic field $h_{x0}\cos(2\pi f t)$ with $h_{x0} = 10$ mT at the middle of the magnet in three cases: (1) $c_j = -323$ m/s, $f = 1$ GHz; (2) $c_j = -808$ m/s, $f = 1$ GHz; and (3) $c_j = -808$ m/s, $f = 5$ GHz. Figure 3 displays the waveforms and the spatial Fourier transformation. In case (1), ferromagnet can accommodate only the RPSWs with wave vectors of $k = -0.052$ and $-0.009$ nm$^{-1}$ (Fig 3 (a) and (d)). In case (2), four degenerate spin waves are identified: two RPSW modes with wave vectors of $k_{+1} = -0.003$ nm$^{-1}$ and $k_{+2} = -0.149$ nm$^{-1}$, and two LPSW modes with wave vectors of $k_{-1} = 0.022$ nm$^{-1}$ and $k_{-2} = 0.129$ nm$^{-1}$ (Fig 3 (b) and (e)). The polarization of spin waves can be confirmed by the phase difference of $m_x$ and $m_y$ component oscillations obtained from the simulation data. For $k_{+1}$ and $k_{+2}$, $m_x$ is always $\pi/2$ ahead of $m_y$, representing RPSWs. While, for $k_{-1}$ and $k_{-2}$, $m_x$ is always $\pi/2$ lagged behind $m_y$, indicating the LPSWs. Figure 3 (g) presents the trajectories of the magnetization precession at different sites in case (2). All the trajectories are ellipses with very large eccentricity, proving that the trajectory is the superposition of a right-handed circular precession and a left-handed circular precession as the superposition trajectory of two right-handed circular precessions with the same frequency is circular. The elliptical trajectory arises from the different amplitudes of the LPSW and RPSW. In addition, the orientation of the trajectory periodically changes with the position due to the difference of wave vectors $k_+$ and $k_-$. The same reason leads to the beat behavior on the waveform (Fig.3 (b). The beat period is $|2\pi / (|k_+| - |k_-|)|$. In the third case, even $|c_j| > c_{j2}$, but $f > f_{\text{-max}} = 3.5$ GHz (see Fig.1), only two RPSW modes with $k = -0.176$ and $0.024$ nm$^{-1}$ are excited (Fig 3 (c) and (f)). The simulated wave vectors have the same values as the ones calculated from the dispersion relations of Eqs. (4) and (5).

The LPSW or RPSW can be excited uniquely by using a spatial exciting field $h_x = h_{x0}\cos(2\pi f t - ky)$ with $k = k_-$ or $k_+$. The waveforms for $m_x$ and $m_y$ components at $t = 50$ ns are shown in Fig. 4(a) when $f = 1$ GHz and $k = k_- = 0.022$ nm$^{-1}$. It shows that the phase of $m_x$ is $\pi/2$ lagged behind $m_y$, confirming the left-handed circular polarization. The trajectory of the magnetization at a given position also demonstrates the left-handed circular precession (Fig. 4 (b)). When $k = k_+ = -0.003$ nm$^{-1}$, the excited spin wave is RPSW as indicated by Fig. 4(c) and (d).

Now we need to test the stability of the spin waves in the presence of the spin-polarized current. Figure 3 (a), (b) and (c) show that spin waves excited by a harmonic field attenuate with propagation distance independent on the propagation direction, indicating the excited spin waves are stable when $|c_j| < c_{j1}$. The attenuation rate for spin waves propagating in the positive $y$-axis is clearly less than that of spin waves propagating in the negative $y$-axis, showing that the spin-polarized current weakens (or enhances) the attenuation of spin waves which propagats in the opposite ( or the same) direction of current[25]. To further confirm the stability of system when $|c_j| < c_{j1}$, we perform the following simulations. Initially, all magnetizations are set randomly deflected 0 ~ 10 degrees from the uniform state, and then the spin-polarized current is applied. All the parameters are the same as those used in the previous simulations. Figure 5 shows the time dependence of the average oscillation amplitude $\langle\sqrt{m_x^2 + m_y^2}\rangle$. It can be seen that when $c_{j2} < |c_j| = 808$ m/s $< c_{j1}$, the system is stable, indicating the LPSWs can stabilize in this current range. If $|c_j| = 1615$ m/s $> c_{j1}$, the system loses its stability.

Finally we discuss the feasibility of generating the LPSWs in experiments. The critical current coefficient $c_{j2}$ for the LPSW can be rewritten as $c_{j2} = 2\sqrt{\gamma q \omega_0}$. Here $\omega_0 = \gamma H$ is the ferromagnetic resonance frequency. Using typical experimental value $\omega_0$ ~ 10 GHz and $q$ ~ $10^{-11}$ A·m, one finds $c_{j2}$ ~ 300 m/s ($c_{j2} = 432$ m/s based on the parameters used in our simulations). The corresponding critical current density $j_2$ is on the order of $10^{13}$ A/m$^2$ calculated from the formula $j = 2eM_s c_j / (g\mu_B P)$ supposing $P = 0.4$ and $M_s$ ~ $10^6$ A/m, which is really large for experiment. But this critical density can be reduced significantly in systems in which Dzyaloshinskii-Moriya interaction (DMI) is present. The DMI introduces a linear term in the dispersion similar to the current[38], and the critical current coefficient for inducing LCP spin waves becomes $c_{j2} = \gamma(-2D/(\mu_0 M_s) \pm 2\sqrt{qH})$ (see Supplementary material for details). Here $D$ is DMI parameter. The sign $\pm$ represents the current flows in the positive (+) or negative (−) $y$-axis direction. Taking $D = 0.8$ mJ/m$^2$, one can find that the critical coefficient $c_{j2}$ is reduced from 300 to 19 m/s and the corresponding current density $j_2$ is $\sim 10^{12}$ A/m$^2$. This current density is accessible experimentally. It should be noted that the DMI can not induce the LPSWs solely. In the absence of the spin-polarized current, two conditions Re$\omega_{\max} \geq 0$ and Im$\omega \leq 0$ are met at the same time, i.e., the system becomes unstable when LPSWs appear.

To conclude, we study theoretically the spin waves in ferromagnetic systems in the presence of spin-polarized electrical current. The systems may keep stable even the current shifts the dispersion curves across the horizontal ordinate. In this case, a new type of spin wave mode, the LPSW is observed together with the RPSW. The nonadiabatic STT plays a key role for the emergence of the LPSWs which can exist only at condition $0 < \beta < 2\alpha$. The micromagnetic simulations are performed to test the stability and the polarization of spin waves. Four-degenerate spin waves, two RPSWs and two LPSWs, are excited simultaneously, confirming the existence of stable LPSWs. The findings in this work may provide a way to use the polarization of spin waves to design new magnonic devices, similar to the photon polarization in photonics.

## ACKNOWLEDGMENT

This work was supported by the National Natural Science Foundation of China Grant Numbers 11674400, 11374373, 11704415. Z.-M.Z. acknowledges support from National Natural Science Foundation of China Grant Number 11474311.

**Figure Captions**

FIG. 1. Dispersion relations of spin waves at different current. The curves 1, 2 and 3 correspond to the $c_j$ = 0, -323 and -808 m/s, respectively. The dotted lines represent the analytical results calculated from Eq. (5) and (6). $f = \omega/2\pi$.

FIG. 2. Phase diagram for the RPSW and LPSW and the spin wave stability.

FIG. 3. Waveforms and the spatial Fourier transformation at $t$ = 50 ns. (a) and (d) for $c_j$ = -323 m/s, $f$ = 1 GHz; (b) and (e) for $c_j$ = -808 m/s, $f$ = 1 GHz; (c) and (f) for $c_j$ = -808 m/s, $f$ = 5 GHz. (g) presents the trajectories of the magnetization precession at different sites when $c_j$ = -808 m/s and $f$ = 1 GHz.

FIG. 4. Waveforms for $m_x$ and $m_y$ components at $t$ = 50 ns excited by field $h_x = h_{x0}\cos(2\pi ft - ky)$ with $h_{x0}$ = 0.05 mT, $f$ = 1 GHz, and $k = k_-$ = 0.022 nm$^{-1}$ (a), $k = k_+$ = -0.003 nm$^{-1}$ (c). The precession trajectories of the magnetization at $y$ = 5000 nm when $k = k_-$ = 0.022 nm$^{-1}$ (b) and $k = k_+$ = -0.003 nm$^{-1}$ (d).

FIG. 5. Time dependence of the average oscillation amplitude of magnetization.

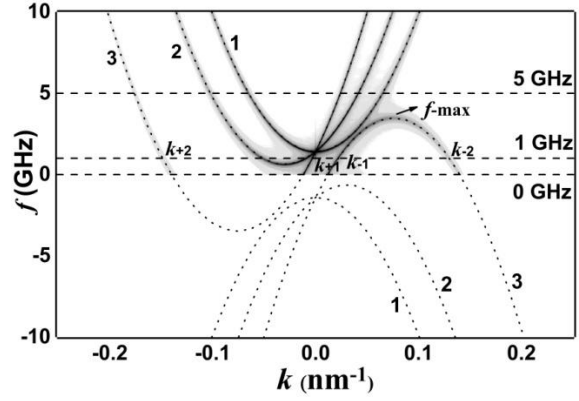

FIG. 1. Dispersion relations of spin waves at different current. The curves 1, 2 and 3 correspond to the $c_j$ = 0, -323 and -808 m/s, respectively. The dotted lines represent the analytical results calculated from Eq. (5) and (6). $f = \omega/2\pi$.

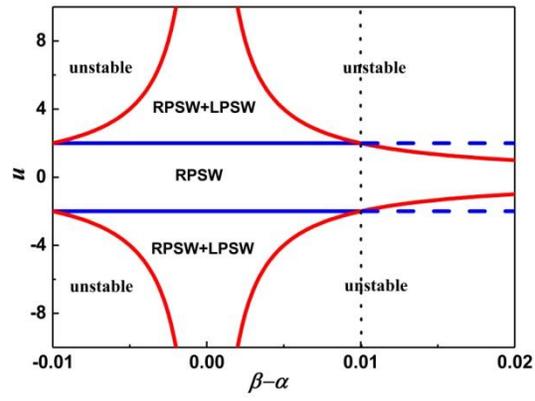

FIG. 2. Phase diagram for the RPSW and LPSW and the spin wave stability.

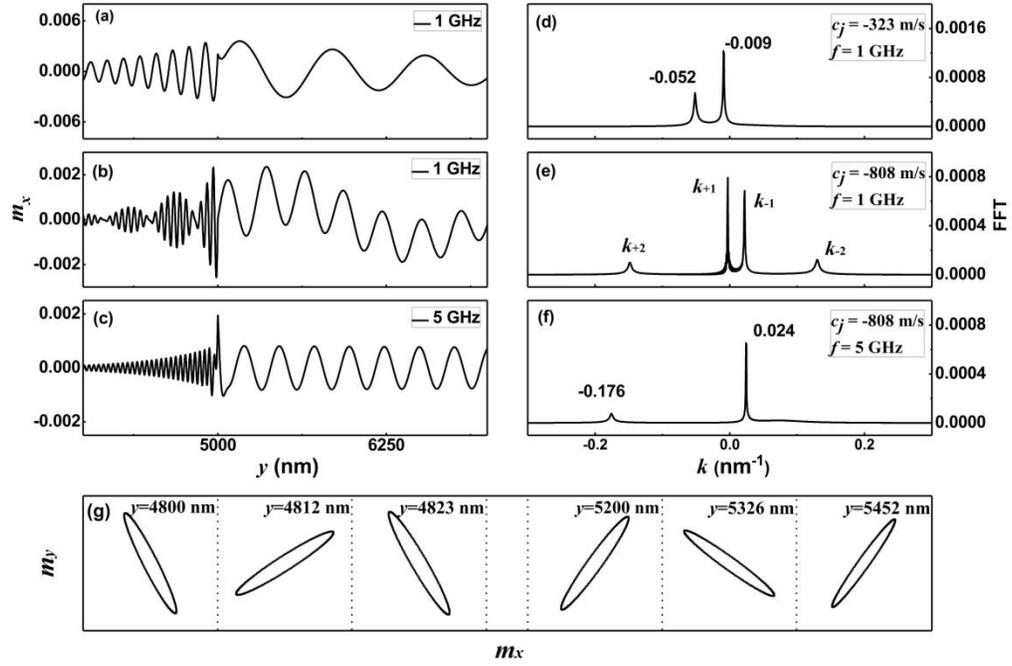

FIG. 3. Waveforms and the spatial Fourier transformation at $t = 50$ ns. (a) and (d) for $c_j = -323$ m/s, $f = 1$ GHz; (b) and (e) for $c_j = -808$ m/s, $f = 1$ GHz; (c) and (f) for $c_j = -808$ m/s, $f = 5$ GHz. (g) presents the trajectories of the magnetization precession at different sites when $c_j = -808$ m/s and $f = 1$ GHz.

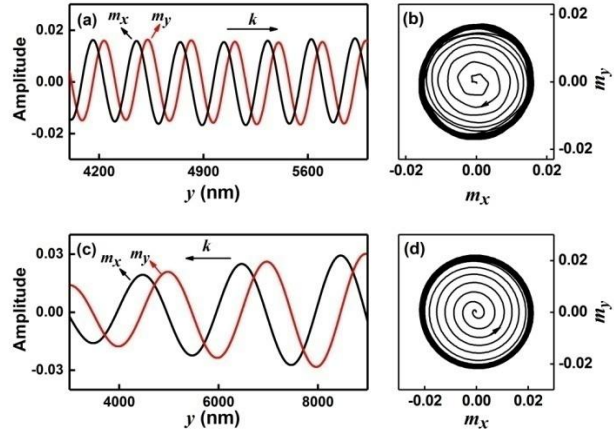

FIG. 4. Waveforms for $m_x$ and $m_y$ components at $t = 50$ ns excited by field $h_x = h_{x0}\cos(2\pi ft - ky)$ with $h_{x0} = 0.05$ mT, $f = 1$ GHz, and $k = k_- = 0.022$ nm$^{-1}$ (a), $k = k_+ = -0.003$ nm$^{-1}$ (c). The precession trajectories of the magnetization at $y = 5000$ nm when $k = k_- = 0.022$ nm$^{-1}$ (b) and $k = k_+ = -0.003$ nm$^{-1}$ (d).

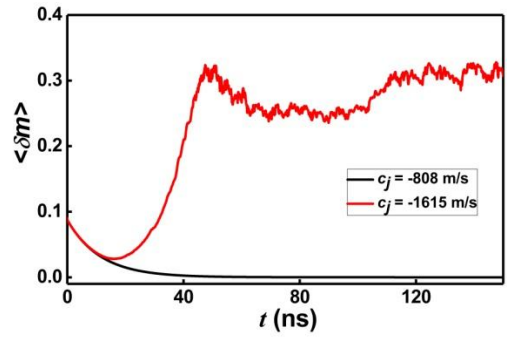

FIG. 5. Time dependence of the average oscillation amplitude of magnetization.